\documentclass[11pt,twoside]{article}
\usepackage{asp2010}

\resetcounters

\begin{document}

\title{Prospects for Asymmetric PNe with ALMA}
\author{P.~J.~Huggins}
\affil{Physics Department, New York University, New York NY 10003, USA}


\begin{abstract}
Millimeter and sub-millimeter observations have made fundamental
contributions to our current understanding of the transition from AGB
stars to white dwarfs. The approaching era of ALMA brings
significantly enhanced observing capabilities at these wavelengths
and promises to push back the frontiers in a number of ways. We
examine the scientific prospects of this new era for PNe, with an
emphasis on how developments may contribute to the goals of the
asymmetric PNe community.
\end{abstract}

\section{Introduction}
The Atacama Large Millimeter/sub-millimeter Array (ALMA), is a new,
major, international telescope that is currently being built in
northern Chile. It will provide significantly enhanced observing
capabilities over existing instrumentation in the mm and sub-mm
wavebands, and is expected to make important contributions to many
areas of astronomy. ALMA's ability to make high quality, high
resolution images in lines and the continuum will provide a new tool
to probe the structure and origin of asymmetric planetary nebulae
(APNe). This paper outlines prospects for APNe with ALMA. Sect.~2
reviews the contributions of mm and sub-mm observations to our current
understanding of the field; Sect.~3 describes ALMA's main
characteristics; and Sect.~4 considers what it might do for APNe.

\section{APNe at mm and sub-mm Wavelengths}

The most important contributions of the mm and sub-mm wavebands to our
current understanding of APNe have been made using molecular line
observations. The low lying rotational transitions of CO, $J=1$--0 at
2.6~mm (115~GHz), $J=2$--1 at 1.3~mm (230~GHz), and $J=3$--2 at
0.87~mm (345~GHz), have been especially useful in probing the
kinematics, distribution, and mass of the molecular gas. Numerous
lines of other molecular species have been detected in spectral scans
of some well-studied objects, e.g., AFGL~618 \citep{pardo07} and
NGC~7027 \citep{zhang08} and these lines provide valuable diagnostics
of physical and chemical conditions.  There is an interesting,
evolving chemistry in the AGB-PN transition \citep{bachiller97}, and
the atomic fine structure lines of C\,{\sc I}, which are useful probes
of photo-dissociation regions, are detectable in the sub-mm
\citep{bachiller94,young97}.

The history of APNe observations in the mm and sub-mm can be divided
into two phases.  In the first phase, the observations were made using
single-dish telescopes. The angular resolution for single-dish
observations is set by the diffraction of the telescope ($\sim
\lambda/D$, where $D$ is the diameter and $\lambda$ the wavelength)
typically $\sim$10--60\hbox{$^{\prime\prime}$}; the unit of
information is the spectrum; and spatial information is obtained by
moving the telescope.  One basic contribution of the single-dish work
has been in setting the baseline for mass-loss on the AGB.  Early
detections \citep[e.g.,][]{knapp82} spawned numerous surveys, so that
we now know the expansion velocity and mass-loss rates of many AGB
stars. A second contribution of single-dish observations has helped
develop one of the central ideas of PN formation: the ejection of
neutral gas in the transition from the AGB. The neutral gas can be
traced from the AGB, through pre-PNe, to \emph{bona fide} PNe
\citep[e.g.,][]{huggins96}, and it makes a direct connection to the
ionized nebulae and their asymmetric structures. A good example is the
CO distribution in the Helix nebula \citep{young99}.  A further
contribution of single dish observations has been the discovery of
high velocity wings in the spectra of pre-PNe and young PNe. This has
revealed a new ejection mechanism with a momentum excess which we
associate with the launching of jets \citep{bujarrabal01} but do not
fully understand.

\begin{figure}[!ht]
\begin{center}
\includegraphics[height=5.2cm]{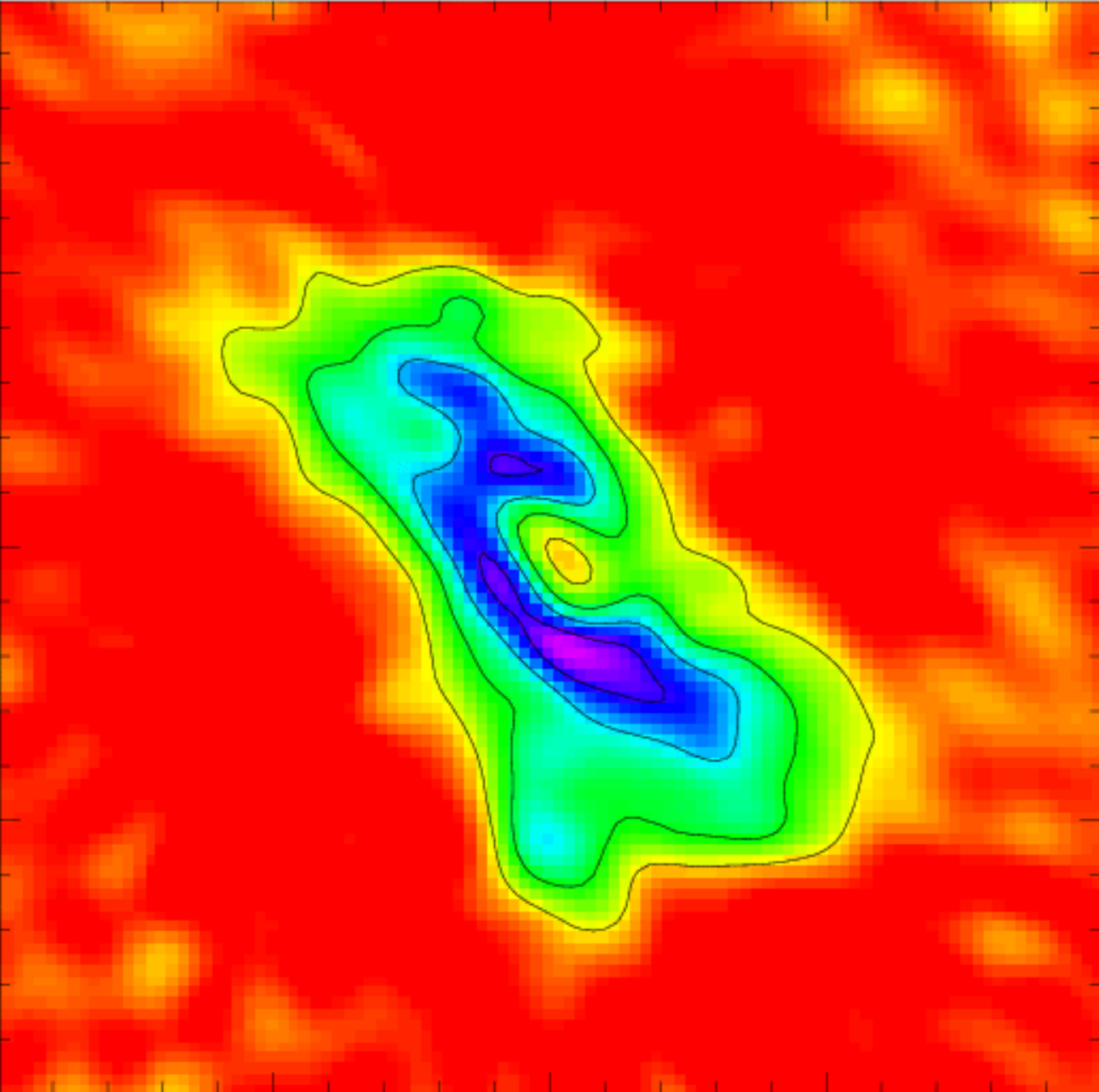} \hspace{0.5cm}
\includegraphics[height=5.2cm]{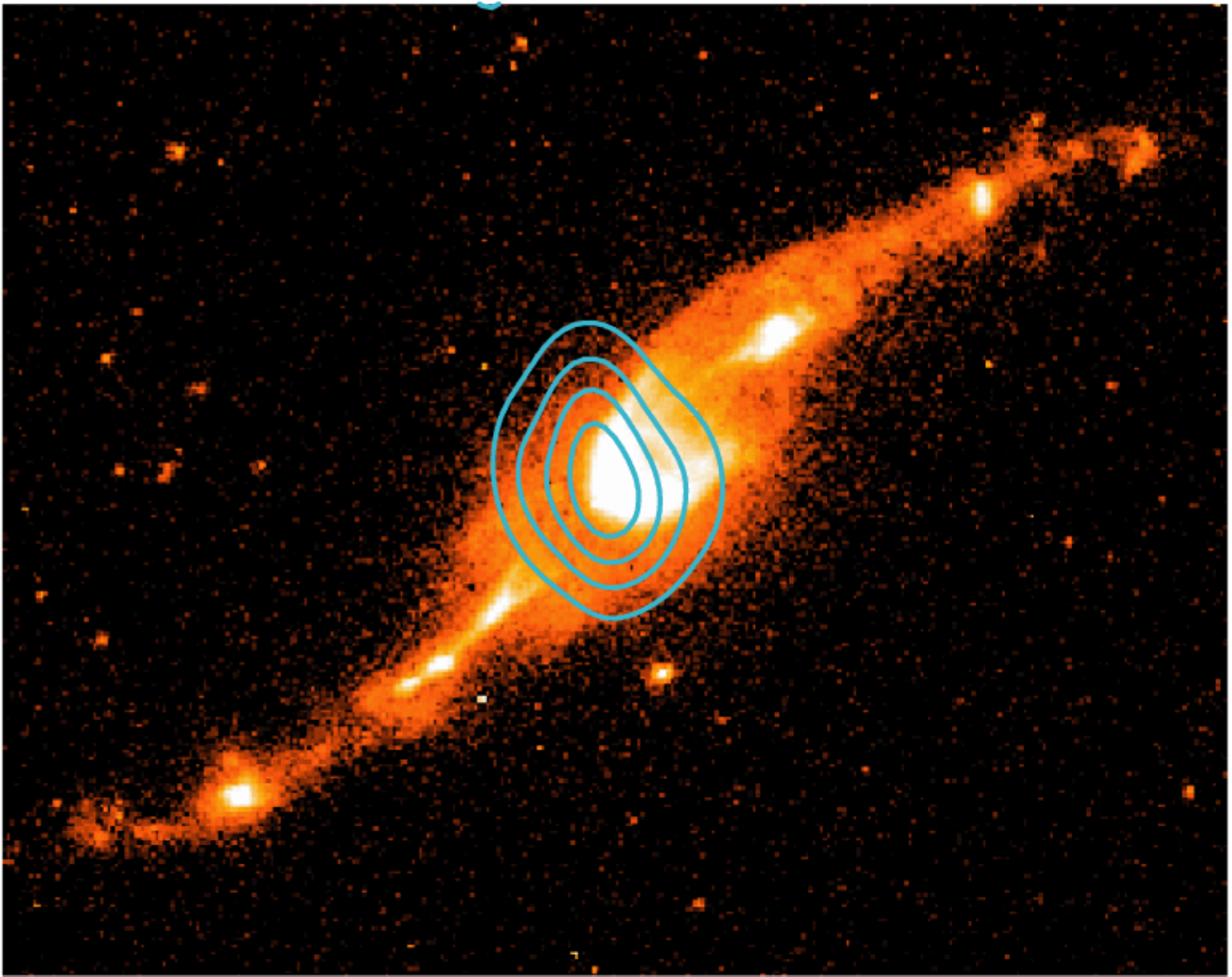}
\caption{Examples of CO line interferometry of APNe. \emph{Left}:
  Image in CO (1--0) integrated intensity of the disk in KjPn~8
  \citep{forveille98}. \emph{Right}: CO (2--1) intensity contours of the
  waist in He~3-1475, superposed on optical HST image \citep{huggins04}.
}
\end{center}
\end{figure}

The second phase of mm and sub-mm observations of APNe has involved
the use of interferometers (notably BIMA, OVRO, SMA, and the IRAM
Plateau de Bure) which provide images/data cubes at higher angular
resolution than single-dish telescopes, from
$\sim$5\hbox{$^{\prime\prime}$} to as high as
$\sim$0.5\hbox{$^{\prime\prime}$}.  CO line interferometry of the
circumstellar envelopes of AGB stars (e.g., \citealt{fong06},
\citealt{castro-carrizo07}, see also Alcolea, this volume) shows that
most (though not all) envelopes are roughly spherically symmetric, but
exhibit multiple arcs and other asymmetries. This sub-structure in the
envelopes is also seen in images of dust-scattered light at optical
wavelengths \citep[e.g.,][]{mauron99,mauron06}.

CO line interferometry of pre-PNe and young PNe, especially in
conjunction with optical HST imaging, has produced striking
results. Two examples, observed with the Plateau de Bure
interferometer in the CO (1--0) and CO (2--1) lines, are shown in
Fig.~1.  Recent examples in other CO lines include: $^{13}$CO (2--1)
in M1-92 \citep{alcolea07}, CO (3--2) in IRAS 22036+5306
\citep{sahai06}, and CO (6--5) in CRL~618 \citep{nakashima07}. These
and similar observations provide a common picture of early PN
formation consisting of enhanced mass-loss in slowly expanding
equatorial tori, with molecular gas entrained in jets along bi-polar
or multi-polar axes.

The results to date reveal a rich phenomenology on size scales of
10$^{16}$--10$^{17}$~cm which is partly explored but not understood.
A number of physical processes may be responsible for the outflows
that produce the asymmetries. For example, a torus could be produced
by rapid rotation, a magnetic explosion, gravitational focusing by a
companion, leakage from Roche lobe overflow, or ejection during a
common envelope phase. However, there is a lack of quantitative
prediction that can clearly discriminate between the various
scenarios. The advent of new opportunities to study these phenomena at
higher resolution with ALMA offers the possibility of important
breakthroughs.

\section{ALMA}

ALMA is a large, imaging interferometer that will operate in the mm
and sub-mm wavebands. It is an international collaboration of North
America, Europe, Japan, and Taiwan, in co-operation with Chile.  The
telescope, which is now being built, will be a significant advance on
current mm and sub-mm instrumentation in terms of angular resolution,
wavelength coverage, and sensitivity. Part of the array is expected to
become available for early science in 2011.

\begin{table}[!ht]
\caption{Characteristics of ALMA}
\smallskip
\begin{center}
{\small
\begin{tabular}{ll}
\tableline
\noalign{\smallskip}
number of antennas  & 50 (12~m) + ACA   \\

frequency range     & 31--950~GHz \\

maximum baselines   & 0.2--16.3~km \ \ $B_\mathrm{rms} = 0.079$--6.6~km \\

primary beam ($\theta_\mathrm{p}$)      &  $21\hbox{$^{\prime\prime}$}\ \times\ 300/
\nu_\mathrm{GHz}$ \\

synthesized beam ($\theta_\mathrm{s}$)    &  $0.08\hbox{$^{\prime\prime}$}\ \times\ 300/
\nu_\mathrm{GHz}\  \times\
1\ \mathrm{km}/B_\mathrm{rms}$ \\

continuum sensitivity$^1$  &  0.10 mJ      \\

line sensitivity$^2$       & 0.10~K ($\theta_\mathrm{s}$ =
1.5\hbox{$^{\prime\prime}$}) --  709~K ($\theta_\mathrm{s}$ = 0.018\hbox{$^{\prime\prime}$}) \\

\noalign{\smallskip}
\tableline
\noalign{\smallskip}
\multicolumn{2}{l}{$^1$ Band 6 (211--275~GHz), t = 60~s \ \ $^2$ Band
  6, t = 60~s, $\Delta
V$ = 1~km~s$^{-1}$  }
\end{tabular}
}
\end{center}
\end{table}

For technical information about ALMA, a useful introduction for
newcomers is
the document: \emph{Observing with ALMA -- A Primer}, by Doug
Johnstone and colleagues. This and other technical information,
including an exposure calculator, can be found at the websites of the
ALMA Regional Centers. Additional technical information, and many
interesting scientific perspectives can be found in the volume:
\emph{Science with the Atacama Millimeter Array: A New Era for
Astrophysics} edited by \citet{bachiller08}.

Some of the principal characteristics of ALMA that determine how it might
be used to observe APNe are listed in Table~1. The main array will
consist of at least 50 $\times$ 12~m antennas, augmented by an
additional small array of 7~m and 12~m antennas -- the Atacama Compact
Array (ACA) -- that can be used for wider field observations, and for
total power measurements.

ALMA will be equipped with an extensive complement of receivers that
will eventually cover almost the entire mm and sub-mm frequency range
from 31 to 950~GHz.  The initial receiver development is for wavebands in the
range 84 to 720~GHz.  The array is located at a high (5,000~m), dry
site to minimize atmospheric effects.  Even so, the atmospheric
transmission at frequencies above about 400~GHz is typically less than
50\%, so this needs to be taken into account in planning observations.

The field of view of ALMA (the primary beam, $\theta_\mathrm{p}$) is
determined by the diffraction of the individual 12~m antennas.
$\theta_\mathrm{p} = 21$\hbox{$^{\prime\prime}$} at 300~GHz and varies
as $\nu^{-1}$. Thus the field of view is very small by the standards
of optical imaging. The field can, of course, be extended by making
additional, adjacent observations (mosaicing) at the expense of
additional observing time.

The angular resolution of ALMA (the synthesized beam,
$\theta_\mathrm{s}$) is determined by the diffraction of the array
(see Table 1). $\theta_\mathrm{s}$ varies as $\nu^{-1}$, and at a
given frequency can be varied by a factor of $\sim$10 according to the
chosen array configuration, which is characterized by the rms baseline
$B_\mathrm{rms}$.  Common setups are likely to have a synthesized beam
of $\sim$0.1\hbox{$^{\prime\prime}$}, so the impact of the instrument
in terms of angular resolution will be somewhat similar to the impact
of the HST at optical wavelengths. At the highest frequencies and most
extended baselines, the resolution is better than
0.01\hbox{$^{\prime\prime}$}, although this is likely to be used only
for special applications.

The resolution of the array is not an independent quantity, because it
is closely linked to the sensitivity of the observations. In measuring
the surface brightness of a spectral line, the noise level (in K)
varies according to the expression:
\[   \Delta T_{rms} \propto  \frac{\theta^{\,2}_\mathrm{p}}{\theta^{\,2}_\mathrm{s}} \  \frac{1}{\eta} \
\frac{1}{\sqrt{N(N-1)}} \ \frac{1}{\sqrt{N_\mathrm{p}}} \
\frac{T_\mathrm{s}}{\sqrt{\Delta \nu\,t} } \] where the first factor on the
right hand side is the
ratio of the primary beam to the synthesized beam, $\eta$ lumps
together various efficiencies, $N$ is the number of antennas, $N_\mathrm{p}$ is
the number of polarizations, and the last factor is the usual
radiometer equation, with $T_\mathrm{s}$ the system temperature, $\Delta\nu$
the observing bandwidth, and $t$ the observing time. Thus the
noise level is proportional to the inverse square of the synthesized
beam. For specificity, some numerical examples for the sensitivity in
Band 6 (211--275~GHz) are given in Table~1, for an observing time of
60~s, and (for a spectral line observation) an effective resolution of
1~km~s$^{-1}$. The back-end of the ALMA system (the correlator) is
extremely flexible and is likely to cover all the spectroscopic
requirements of the APN field. The sensitivity in the continuum is
particularly high because of the wide (8~GHz) bandwidth available.

From the equation above, it can be seen that the large number of
antennas ($N=50$) plays an important role in the sensitivity of the
array. The corresponding number of baselines, given by $N(N-1)/2$, is
even larger (1225).  This means that the $u-v$ (Fourier) plane is well
sampled, and leads to high quality imaging, even in snapshot
(short-exposure) modes.  Overall, the gain in sensitivity compared to
the state-of-the-art Plateau de Bure interferometer (at $\sim$ 230~GHz
for the same synthesized beam, etc.) is a factor of $\sim$15--20. The
images shown in Fig.~1 could in principle be obtained in observing
times $\sim$1~min.  Thus ALMA is indeed a major development in mm and
sub-mm instrumentation.

\section{Strategies and Projects}

There are many ways in which the capabilities of ALMA can be used to
study APNe. Strategies range from using the speed of the instrument to
carry out snapshot surveys, to exploiting the highest resolution modes
to probe details of objects of special interest. Here we outline some
of the possibilities, with an emphasis on the scientific objectives.

\subsection{AGB Stars} \cite{olofsson08} has reviewed the general
prospects for AGB stars with ALMA, and the reader is referred to his
paper for details.  Important developments will be observations of the
dust forming regions, and sensitive studies of the chemistry. For
APNe, one interesting aspect is that the photospheres and close
environs of the nearest AGB stars will be resolved in the continuum
with ALMA in the highest resolution configurations. This is a new type
of probe at these wavelengths and may have an important bearing on
understanding early APN formation. Low luminosity companion stars will
not be directly detected in the glare of the AGB star, but they could
generate detectable regions of ionized gas under some circumstances,
as in the case of Mira at longer wavelengths \citep{matthews06}.

\subsection{AGB Envelopes} As the precursors of APNe,
the circumstellar envelopes of AGB stars provide a number of important
constraints on the origins of PN asymmetries.  First, there is good
evidence from optical HST imaging for incipient core activity in some
AGB envelopes, as discussed by Sahai (this volume). This activity may
be caused by young or weak jets, or some other type of activity which
has not yet fully developed.  Thus it would be important to determine
how widespread this core activity is, and to characterize its
structure and kinematics. This could be done using CO line
observations of envelope cores at high resolution.

A second perspective on AGB envelopes is directly concerned with
probing the presence of a binary companion. If axi-symmetry is induced
by interaction with a companion, the secondary star must have been
present throughout the entire AGB phase and should leave its imprint
on the circumstellar envelope, as emphasized by \citet{huggins09}.

One way the imprint is effected is that the reflex motion of the AGB
star induces a spiral pattern in the circumstellar envelope (e.g.,
\citealt{mastrodemos99}, see also Raga, this volume). A clear example
of the spiral pattern has been seen in the circumstellar envelope of
AFGL~3068 \citep{mauron06,morris06} in dust-scattered light
observations. This pattern may also be detectable in CO or some other
line affected by the weak spiral shock.  The radial wavelength of the
spiral is $\lambda \sim VP$ where $V$ is the expansion velocity of the
envelope, and $P$ is the period of the binary. For AFGL~3068 $P \sim
800$~yr and $V\sim 14$~km~s$^{-1}$. At a distance of $\sim 1$~kpc, the
angular separation of the arms is $\sim
2$\hbox{$^{\prime\prime}$}. Thus the pattern is in principle
relatively easy to resolve at high resolution in other nearby systems
with intermediate or long periods.

A related signature of a companion star is the degree to which its
gravitational field flattens the AGB envelope. This global shaping of
the envelope is the simplest characteristic of a binary companion to
observe on a large scale. The magnitude of the effect, which depends
on the companion mass and separation and the wind velocity, has been
discussed in detail by \citet{huggins09}.  The angular size of an AGB
envelope in CO for a mass-loss rate $\ga$10$^{-5}$~M$_\odot$~yr$^{-1}$ is
$\ga$25\hbox{$^{\prime\prime}$}/D$_{kpc}$ where $D$ is the distance. Thus
envelope shapes can be measured for large numbers of AGB stars to
probe this effect.

\subsection{Magnetic Fields} The role of magnetic fields in AGB
stars and PNe has long been a controversial topic. The problem used to
be magnetic fields versus binary companions for shaping PNe, but now
that companions seem to be fairly common, and their interactions can
generate magnetic fields, the issue has changed focus.  There is
probably a consensus that MHD is needed to launch and form jets
(Frank, this volume). The main issue seems to be: what else do
magnetic fields do?

Radio astronomers have long reported the presence of dynamically
important magnetic fields in maser spots in AGB envelopes (e.g.,
Bains, this volume). However, the spots are small, so there is a
question whether the strong fields are local or global.  \citet{herpin09}
have recently reported magnetic field measurements using mm
CN emission (which is not in spots) at levels equivalent to the fields
in the spots. Hence the fields may be globally important. ALMA's
ability to measure polarization and thereby to map the magnetic field
strength and geometry is likely to be important in sorting out this
issue. See Vlemmings (this volume) for more details.

\subsection{Shaping in Pre-PNe} One of the most direct applications
of ALMA to the problems of APN formation will be in characterizing the
early jet shaping in pre-PNe. As explained in Sect.~2, there is
already a generic picture (at least for a subset of PNe), based on
current interferometry, as shown in the left hand panel of Fig.~2.

\begin{figure}[!ht]
\begin{center}
\includegraphics[height=5.0cm]{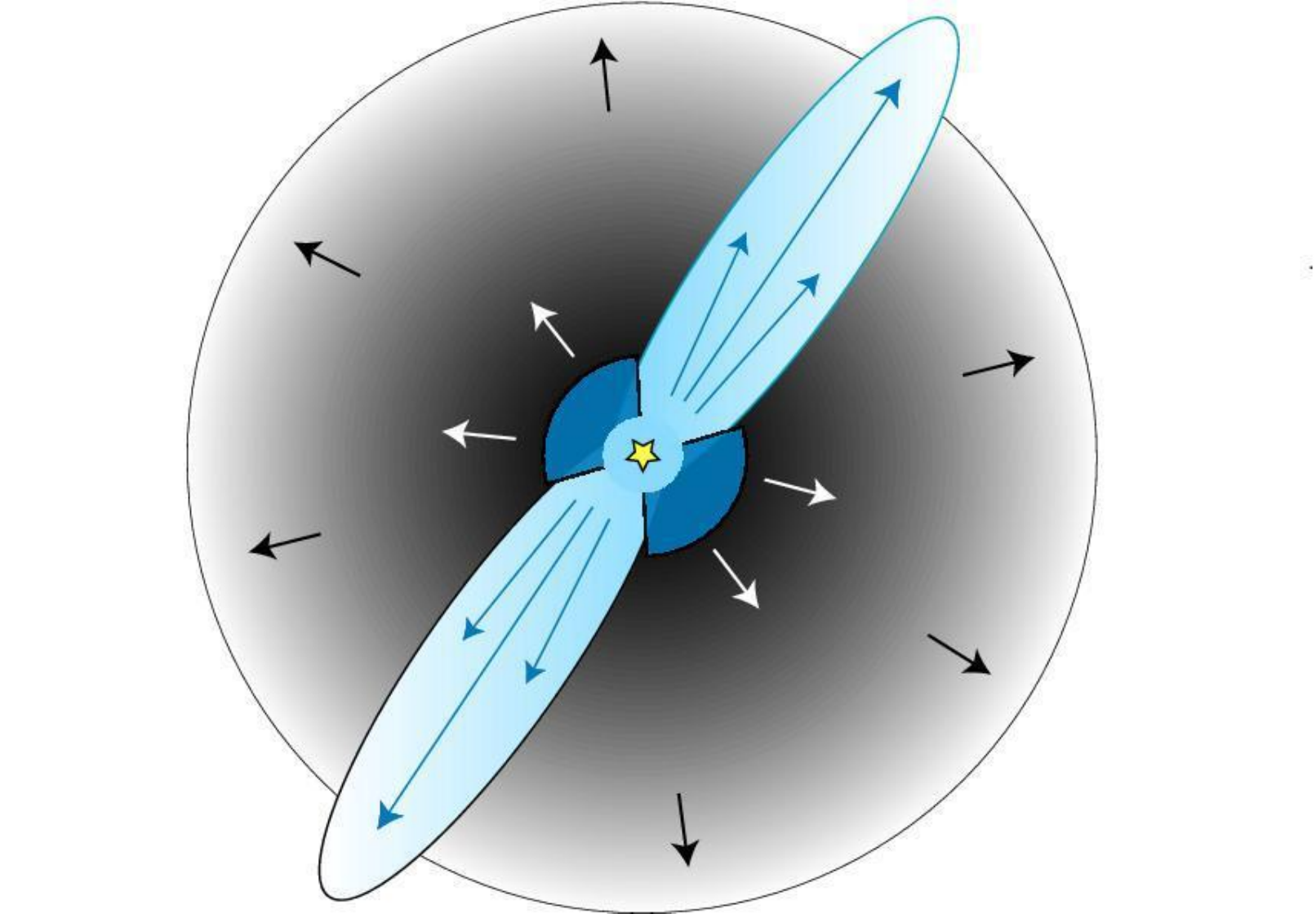}
\includegraphics[height=5.0cm]{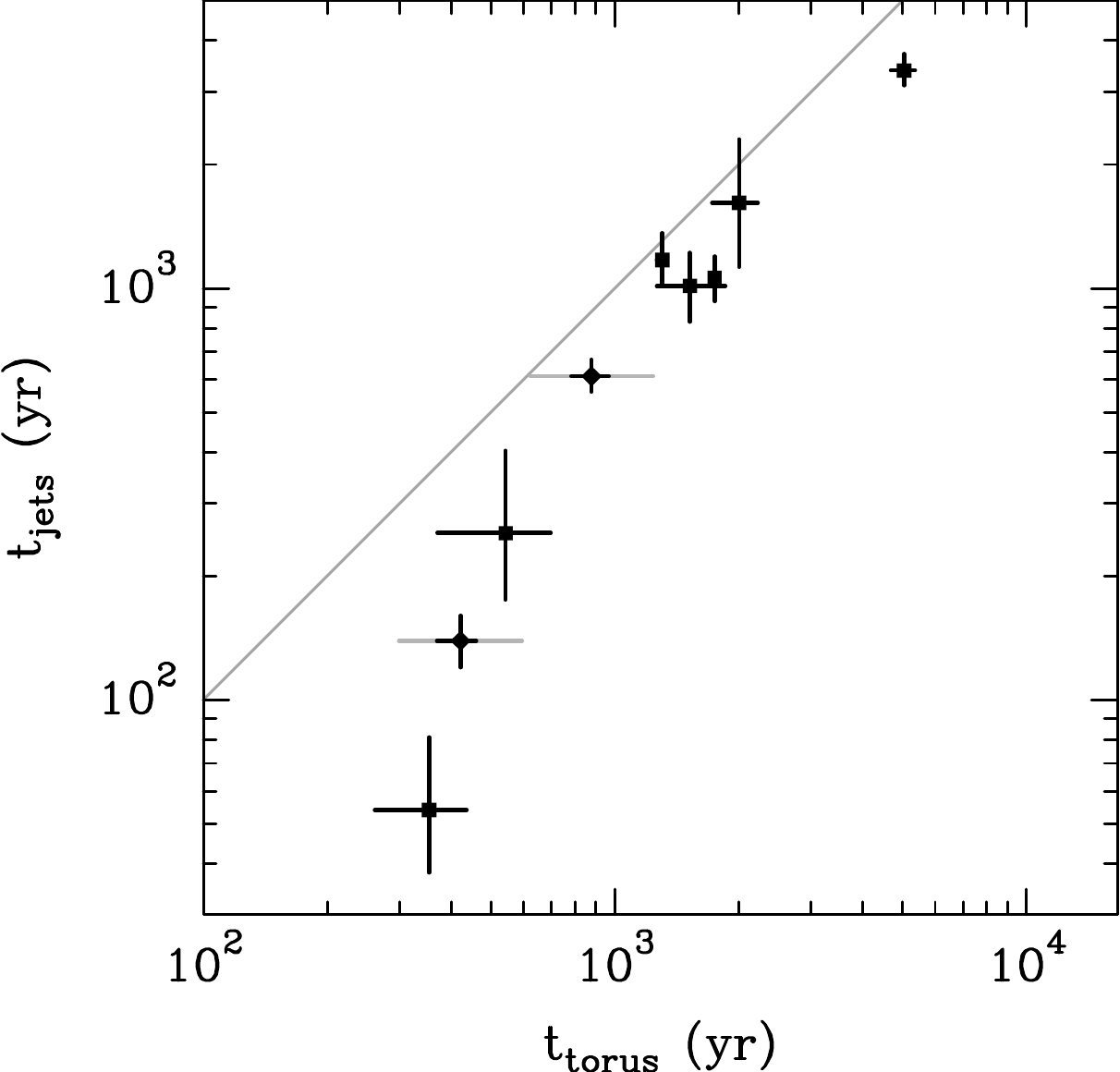} \hspace{0.5cm}
\caption{\emph{Left}: Schematic picture of the early shaping of PNe by
  jets and equatorial outflows. \emph{Right}: Correlation of the
  kinematic ages of jets and equatorial tori \citep[from][]{huggins07}.  }
\end{center}
\end{figure}

This picture is only partly explored by current observations. The
detailed properties accessible to line observations with ALMA include:
the geometry and kinematics of the jets and equatorial outflows, the
energy and momentum budget, and possible connections with extended
circum-binary disks, which are common in other classes of post-AGB
stars. A systematic investigation of these and related properties is
likely to provide important clues on the physics that controls the
outflows.  For example, one result based on current mm interferometry
is that the kinematic ages of jets and tori are correlated
(Fig.~2). They have nearly the same ages, and are therefore ejected
nearly simultaneously \citep{huggins07}. This points to a fundamental
connection between the outflows. It rules out some classes of physical
model, and even suggests that slight timing differences between jets
and tori can probe the properties of unseen accretion disks that are
believed by some to be responsible for launching jets. Much more
detailed information along these lines is expected to be very
productive.

\subsection{Globules and Debris Disks} ALMA also has interesting
applications in fully developed PNe.  One is probing the excess long
wavelength continuum emission from debris disks surrounding the
central stars of some PNe. The first of these was detected using
Spitzer observations of the Helix nebula \citep{su07}. The estimated
size is $\sim$100~AU. Although the dust continuum is expected to be
falling in the sub-mm and mm wavebands, the emission from the nearest
examples is detectable with ALMA and the structure should be resolved.

A second application to \emph{bona fide} PNe is the small scale
structure in the neutral gas -- the globules, as exemplified by those
in the Helix nebula.  The Helix globules are about
1\hbox{$^{\prime\prime}$} in diameter with extended tails, and have
already been partially resolved using mm interferometry in CO
\citep{huggins02}. A full characterization of the internal structure
and kinematics of the globules and their tails is entirely feasible in
the Helix and other nearby PNe, and would be an important step in
refining our understanding of their origins.

\section{Concluding Remarks}

ALMA will be a significant advance on current mm and sub-mm
instrumentation, and it will have considerable impact in many areas of
astronomy. For APNe, there are major unsolved problems, especially
with the physical processes that generate the asymmetries. Several
aspects of these are well suited to ALMA's capabilities. Some
approaches are direct extensions of current interferometric
observations to higher resolution and sensitivity (e.g., CO imaging),
and some are new types of probes (e.g., line polarization). Overall
the prospects are excellent for taking the field of APNe forward to
the next level.

\acknowledgements I thank Pierre Cox for comments on the
manuscript. This work is supported in part by NSF grant AST 08-06910.

\bibliography{huggins}

\begin{thebibliography}{}
\expandafter\ifx\csname natexlab\endcsname\relax\def\natexlab#1{#1}\fi
\expandafter\ifx\csname url\endcsname\relax
  \def\url#1{\texttt{#1}}\fi
\expandafter\ifx\csname urlprefix\endcsname\relax\def\urlprefix{URL }\fi
\providecommand{\eprint}[2][]{\url{#2}}

\bibitem[{{Alcolea} et~al.(2007){Alcolea}, {Neri}, \& {Bujarrabal}}]{alcolea07}
{Alcolea}, J., {Neri}, R., \& {Bujarrabal}, V. 2007, \aap, 468, L41

\bibitem[{{Bachiller} \& {Cernicharo}(2008)}]{bachiller08}
{Bachiller}, R., \& {Cernicharo}, J. (eds.) 2008, Science With the Atacama
  Millimeter Array: A New Era for Astrophysics, vol. 313 of Ap\&SS (New York:
  Springer)

\bibitem[{{Bachiller} et~al.(1997){Bachiller}, {Forveille}, {Huggins}, \&
  {Cox}}]{bachiller97}
{Bachiller}, R., {Forveille}, T., {Huggins}, P.~J., \& {Cox}, P. 1997, \aap,
  324, 1123

\bibitem[{{Bachiller} et~al.(1994){Bachiller}, {Huggins}, {Cox}, \&
  {Forveille}}]{bachiller94}
{Bachiller}, R., {Huggins}, P.~J., {Cox}, P., \& {Forveille}, T. 1994, \aap,
  281, L93

\bibitem[{{Bujarrabal} et~al.(2001){Bujarrabal}, {Castro-Carrizo}, {Alcolea},
  \& {S{\'a}nchez Contreras}}]{bujarrabal01}
{Bujarrabal}, V., {Castro-Carrizo}, A., {Alcolea}, J., \& {S{\'a}nchez
  Contreras}, C. 2001, \aap, 377, 868

\bibitem[{{Castro-Carrizo} et~al.(2007){Castro-Carrizo}, {Neri}, {Winters},
  {Bujarrabal}, {Quintana-Lacaci}, {Alcolea}, {Sch{\"i}er}, {Olofsson}, \&
  {Lindqvist}}]{castro-carrizo07}
{Castro-Carrizo}, A., {Neri}, R., {Winters}, J.~M., {Bujarrabal}, V.,
  {Quintana-Lacaci}, G., {Alcolea}, J., {Sch{\"i}er}, F.~L., {Olofsson}, H., \&
  {Lindqvist}, M. 2007, in Why Galaxies Care About AGB Stars: Their Importance
  as Actors and Probes, edited by {F.~Kerschbaum, C.~Charbonnel, \&
  R.~F.~Wing}, vol. 378 of ASP Conf. Ser., 199

\bibitem[{{Fong} et~al.(2006){Fong}, {Meixner}, {Sutton}, {Zalucha}, \&
  {Welch}}]{fong06}
{Fong}, D., {Meixner}, M., {Sutton}, E.~C., {Zalucha}, A., \& {Welch}, W.~J.
  2006, \apj, 652, 1626

\bibitem[{{Forveille} et~al.(1998){Forveille}, {Huggins}, {Bachiller}, \&
  {Cox}}]{forveille98}
{Forveille}, T., {Huggins}, P.~J., {Bachiller}, R., \& {Cox}, P. 1998, \apjl,
  495, L111

\bibitem[{{Herpin} et~al.(2009){Herpin}, {Baudy}, {Josselin}, {Thum}, \&
  {Wiesemeyer}}]{herpin09}
{Herpin}, F., {Baudy}, A., {Josselin}, E., {Thum}, C., \& {Wiesemeyer}, H.
  2009, in IAU Symposium, vol. 259 of IAU Symposium, 47

\bibitem[{{Huggins}(2007)}]{huggins07}
{Huggins}, P.~J. 2007, \apj, 663, 342

\bibitem[{{Huggins} et~al.(1996){Huggins}, {Bachiller}, {Cox}, \&
  {Forveille}}]{huggins96}
{Huggins}, P.~J., {Bachiller}, R., {Cox}, P., \& {Forveille}, T. 1996, \aap,
  315, 284

\bibitem[{{Huggins} et~al.(2002){Huggins}, {Forveille}, {Bachiller}, {Cox},
  {Ageorges}, \& {Walsh}}]{huggins02}
{Huggins}, P.~J., {Forveille}, T., {Bachiller}, R., {Cox}, P., {Ageorges}, N.,
  \& {Walsh}, J.~R. 2002, \apjl, 573, L55

\bibitem[{{Huggins} et~al.(2009){Huggins}, {Mauron}, \& {Wirth}}]{huggins09}
{Huggins}, P.~J., {Mauron}, N., \& {Wirth}, E.~A. 2009, \mnras, 396, 1805

\bibitem[{{Huggins} et~al.(2004){Huggins}, {Muthu}, {Bachiller}, {Forveille},
  \& {Cox}}]{huggins04}
{Huggins}, P.~J., {Muthu}, C., {Bachiller}, R., {Forveille}, T., \& {Cox}, P.
  2004, \aap, 414, 581

\bibitem[{{Knapp} et~al.(1982){Knapp}, {Phillips}, {Leighton}, {Lo}, {Wannier},
  {Wootten}, \& {Huggins}}]{knapp82}
{Knapp}, G.~R., {Phillips}, T.~G., {Leighton}, R.~B., {Lo}, K.~Y., {Wannier},
  P.~G., {Wootten}, H.~A., \& {Huggins}, P.~J. 1982, \apj, 252, 616

\bibitem[{{Mastrodemos} \& {Morris}(1999)}]{mastrodemos99}
{Mastrodemos}, N., \& {Morris}, M. 1999, \apj, 523, 357

\bibitem[{{Matthews} \& {Karovska}(2006)}]{matthews06}
{Matthews}, L.~D., \& {Karovska}, M. 2006, \apjl, 637, L49

\bibitem[{{Mauron} \& {Huggins}(1999)}]{mauron99}
{Mauron}, N., \& {Huggins}, P.~J. 1999, \aap, 349, 203

\bibitem[{{Mauron} \& {Huggins}(2006)}]{mauron06}
--- 2006, \aap, 452, 257

\bibitem[{{Morris} et~al.(2006){Morris}, {Sahai}, {Matthews}, {Cheng}, {Lu},
  {Claussen}, \& {S{\'a}nchez-Contreras}}]{morris06}
{Morris}, M., {Sahai}, R., {Matthews}, K., {Cheng}, J., {Lu}, J., {Claussen},
  M., \& {S{\'a}nchez-Contreras}, C. 2006, in Planetary Nebulae in our Galaxy
  and Beyond, edited by {M.~J.~Barlow \& R.~H.~M{\'e}ndez}, vol. 234 of IAU
  Symposium, 469

\bibitem[{{Nakashima} et~al.(2007){Nakashima}, {Fong}, {Hasegawa}, {Hirano},
  {Koning}, {Kwok}, {Lim}, {Dinh-Van-Trung}, \& {Young}}]{nakashima07}
{Nakashima}, J., {Fong}, D., {Hasegawa}, T., {Hirano}, N., {Koning}, N.,
  {Kwok}, S., {Lim}, J., {Dinh-Van-Trung}, \& {Young}, K. 2007, \aj, 134, 2035

\bibitem[{{Olofsson}(2008)}]{olofsson08}
{Olofsson}, H. 2008, \apss, 313, 201

\bibitem[{{Pardo} et~al.(2007){Pardo}, {Cernicharo}, {Goicoechea},
  {Gu{\'e}lin}, \& {Asensio Ramos}}]{pardo07}
{Pardo}, J.~R., {Cernicharo}, J., {Goicoechea}, J.~R., {Gu{\'e}lin}, M., \&
  {Asensio Ramos}, A. 2007, \apj, 661, 250

\bibitem[{{Sahai} et~al.(2006){Sahai}, {Young}, {Patel}, {S{\'a}nchez
  Contreras}, \& {Morris}}]{sahai06}
{Sahai}, R., {Young}, K., {Patel}, N.~A., {S{\'a}nchez Contreras}, C., \&
  {Morris}, M. 2006, \apj, 653, 1241

\bibitem[{{Su} et~al.(2007){Su}, {Chu}, {Rieke}, {Huggins}, {Gruendl},
  {Napiwotzki}, {Rauch}, {Latter}, \& {Volk}}]{su07}
{Su}, K.~Y.~L., {Chu}, Y., {Rieke}, G.~H., {Huggins}, P.~J., {Gruendl}, R.,
  {Napiwotzki}, R., {Rauch}, T., {Latter}, W.~B., \& {Volk}, K. 2007, \apjl,
  657, L41

\bibitem[{{Young} et~al.(1997){Young}, {Cox}, {Huggins}, {Forveille}, \&
  {Bachiller}}]{young97}
{Young}, K., {Cox}, P., {Huggins}, P.~J., {Forveille}, T., \& {Bachiller}, R.
  1997, \apjl, 482, L101

\bibitem[{{Young} et~al.(1999){Young}, {Cox}, {Huggins}, {Forveille}, \&
  {Bachiller}}]{young99}
--- 1999, \apj, 522, 387

\bibitem[{{Zhang} et~al.(2008){Zhang}, {Kwok}, \& {Dinh-V-Trung}}]{zhang08}
{Zhang}, Y., {Kwok}, S., \& {Dinh-V-Trung} 2008, \apj, 678, 328

\end{thebibliography}

\end{document}